\documentclass[aps,pre,superscriptaddress,floatfix]{revtex4-2}

\usepackage[colorlinks,bookmarks=false,citecolor=blue,linkcolor=blue,urlcolor=blue]{hyperref}
\usepackage[all]{hypcap}   

\usepackage{amsmath,amssymb}
\usepackage{graphicx}

\usepackage{verbatim}
\usepackage{color}

\usepackage{placeins}    
\usepackage{flafter}     
\usepackage{color}
\usepackage{bbold}

\usepackage[normalem]{ulem}

\usepackage[capitalise]{cleveref}
\crefname{appendix}{}{}
\crefname{section}{Sec.}{}

\newcommand{\ue}{\text{e}}
\newcommand{\ui}{\text{i}}
\newcommand{\ud}{\text{d}}
\newcommand{\uL}{L}
\newcommand{\uR}{R}

\newcommand{\gnat}{\ensuremath{\gamma_{\mathrm{nat}}}}
\newcommand{\gtyp}{\ensuremath{\gamma_{\mathrm{typ}}}}
\newcommand{\ginv}{\ensuremath{\gamma_{\mathrm{inv}}}}

\newcommand{\Imag}{\ensuremath{\mathrm{Im}}}
\newcommand{\Real}{\ensuremath{\mathrm{Re}}}
\newcommand{\vecx}{\ensuremath{\boldsymbol{x}}}
\newcommand{\vecy}{\ensuremath{\boldsymbol{y}}}
\newcommand{\vecr}{\ensuremath{\boldsymbol{r}}}
\newcommand{\Hus}{\ensuremath{\mathcal{H}}}

\newcommand{\HusR}{\ensuremath{\Hus^\mathrm{R}}}
\newcommand{\HusL}{\ensuremath{\Hus^\mathrm{L}}}
\newcommand{\HusLR}{\ensuremath{\Hus^\mathrm{LR}}}
\newcommand{\HusLRAvg}{\ensuremath{\langle \HusLR \rangle_\gamma}}

\newcommand{\hLR}{\ensuremath{h^\mathrm{LR}}}

\newcommand{\psiR}{\ensuremath{\psi^\mathrm{R}}}
\newcommand{\psiL}{\ensuremath{\psi^\mathrm{L}}}
\newcommand{\etaR}{\ensuremath{\eta^\mathrm{R}}}
\newcommand{\etaL}{\ensuremath{\eta^\mathrm{L}}}
\newcommand{\etaLR}{\ensuremath{\eta^\mathrm{LR}}}

\newcommand{\muR}{\ensuremath{\mu_\gamma^\mathrm{R}}}
\newcommand{\muL}{\ensuremath{\mu_\gamma^\mathrm{L}}}
\newcommand{\muLR}{\ensuremath{\mu_\gamma^\mathrm{LR}}}

\newcommand{\rhoR}{\ensuremath{\varrho_\gamma^\mathrm{R}}}
\newcommand{\rhoL}{\ensuremath{\varrho_\gamma^\mathrm{L}}}
\newcommand{\rhoLR}{\ensuremath{\varrho_\gamma^\mathrm{LR}}}
\newcommand{\PLR}{\ensuremath{P^\mathrm{LR}}}

\newcommand{\Rcav}{\ensuremath{R_\mathrm{cav}}}
\newcommand{\Resc}{\ensuremath{R_\mathrm{refl}}}

\makeatletter
\let\Hy@backout\@gobble
\makeatother
\begin{document}

\title{
Left-Right Husimi Representation of Chaotic Resonance States: \\
Invariance and Factorization
}

\author{Florian Lorenz}
\thanks{These authors contributed equally to this work.}
\affiliation{TU Dresden,
 Institute of Theoretical Physics and Center for Dynamics,
 01062 Dresden, Germany}

\author{Jan Möseritz-Schmidt}
\thanks{These authors contributed equally to this work.}
\affiliation{TU Dresden,
 Institute of Theoretical Physics and Center for Dynamics,
 01062 Dresden, Germany}

\author{Roland Ketzmerick}
\affiliation{TU Dresden,
 Institute of Theoretical Physics and Center for Dynamics,
 01062 Dresden, Germany}

\date{\today}

\begin{abstract}
    For chaotic scattering systems we investigate the left-right Husimi
    representation, which combines left and right resonance states.
    We demonstrate that the left-right Husimi representation is invariant in the
    semiclassical limit under the corresponding closed classical dynamics, which
    we call quantum invariance.
    Furthermore, we show that it factorizes into a classical multifractal
    structure times universal quantum fluctuations.
    Numerical results for a dielectric cavity, the three-disk scattering system,
    and quantum maps confirm both the quantum invariance and the factorization.
\end{abstract}

\maketitle

\section{Introduction}

Understanding the structure of resonance states in systems with a chaotic
classical phase space is an essential goal in quantum chaos~\cite{BorGuaShe1991,
Gas1998, CasMasShe1999b, LeeRimRyuKwoChoKim2004, SchTwo2004,
KeaNovPraSie2006, NonRub2007, KeaNonNovSie2008, ErmCarSar2009,
AltPorTel2013, BilGarGeoGir2019, HalMalGra2023, ReySigPraSan2024,
YiDieHanRyu2025}.
Much progress has been made on right resonance states, which
are conjectured to
be the product of a factor of classical origin giving a multifractal structure
depending on the decay rate times a universal factor given by exponentially
distributed quantum fluctuations~\cite{ClaKunBaeKet2021, KetClaFriBae2022,
SchKet2023, KetLorSch2025}.
This factorization conjecture for right resonance states was studied
in position space and in phase space using the Husimi representation.
Recently, for the factor of classical origin a conditionally invariant measure
with given decay rate was
constructed based on a generalization of Ulam's method~\cite{KetLorSch2025},
replacing previous approximate methods~\cite{ClaKoeBaeKet2018, ClaAltBaeKet2019}.
In Ref.~\cite{KetLorSch2025} it is applied to systems with full and partial
escape~\cite{AltPorTel2013},
e.g.\ the three-disk scattering system (full escape)~\cite{GasRic1989c,
CviEck1989, WeiBarKuhPolSch2014} and dielectric cavities (partial
escape)~\cite{CaoWie2015}.

There are also alternative representations of resonance states that are useful
to reveal their properties.
One such representation was introduced by Ermann, Carlo, and
Saraceno~\cite{ErmCarSar2009}, which we call left-right Husimi representation
(LR-Husimi representation).
This representation combines left and right resonance states and is naturally
related to the spectral decomposition of the time-evolution operator of the open
system.
The LR-Husimi representation was introduced for systems with full escape to
investigate the quantum analogue of the classical chaotic saddle.
In this representation it was found that scarring on short periodic orbits
is especially prominent~\cite{ErmCarSar2009}.
A constructive semiclassical approach allows for describing resonance states
in the LR-Husimi representation using scar functions on short periodic
orbits~\cite{NovPedWisCarKea2009, ErmCarPedSar2012, PedWisCarNov2012,
CarWisErmBenBor2013}.
This was later extended to systems with partial escape~\cite{CarBenBor2016,
PraCarBenBor2018} and localization properties were studied~\cite{MonCarBor2024,
MonErmRivBorCar2025}.
Also from a mathematical perspective there is an interest in the LR-Husimi
representation~\cite{WeiBarKuhPolSch2014, BarSchWei2022, SchWeiBar2023}.
For uniformly hyperbolic system with full escape it was shown that individual
resonance states in LR-Husimi representation can be described by Ruelle
resonances of zeta functions.

It is an open question whether the LR-Husimi representation can be factorized in
a similar way as right resonance states, namely into a factor of classical
origin times universal fluctuations.
If so, what is the interpretation of the factor of classical origin?
For systems with full escape it should be related to the invariant chaotic
saddle, while for systems with partial escape the interpretation is open.
The fluctuations were studied in Ref.~\cite{MonCarBor2024} and significant
deviations from an exponential distribution were found, which were interpreted
as a localization effect.
Instead, we expect that exponential fluctuations from left and right resonance
states have to be combined.
Are they correlated, what is the resulting distribution, and is it universal?

A further set of questions is related to a possible invariance of the LR-Husimi
representation.
For full escape this might be expected, as the LR-Husimi representation is
supported by the invariant chaotic saddle.
If so, is the LR-Husimi representation also invariant for systems with partial
escape?
Is it invariant under the classical dynamics with or without escape,
i.e.\ the open or closed dynamics?
Is this invariance exact or approximate and what happens in the semiclassical
limit?

In this paper we show that the LR-Husimi representation of resonance states in
classically chaotic systems with full and partial escape has two important
properties:
(i) invariance under the closed dynamics in the semiclassical limit, which we
call quantum invariance;
(ii) factorization with one factor of classical origin times universal quantum
fluctuations.
We derive the distribution of the fluctuations, based on the assumption of
independent right and left fluctuations.
We find that the factor of classical origin is obtained by combining the
conditionally invariant measures corresponding to left and right resonance
states.
Numerical support for properties (i) and (ii) is given for a dielectric cavity
(partial escape), the three-disk scattering system (full escape), and quantum
maps with partial and full escape.

The paper is organized as follows:
In \cref{sec:phase_space_representation} we review the definition of left and
right resonance states and their LR-Husimi representation for open quantum maps
and time-continuous scattering systems.
In \cref{sec:invariance} we show that the LR-Husimi representation is
invariant under the closed dynamics in the semiclassical limit.
In \cref{sec:factorization} we derive that the LR-Husimi representation can be
factorized into a product of a factor of classical origin times universal
fluctuations, which is numerically verified in
\cref{sec:numerical_verification}.
Finally, we summarize our results and outline further directions in
\cref{sec:discussion}.

\section{Phase space representation of resonance states}
\label{sec:phase_space_representation}

We review the definition of left and right resonance states and their LR-Husimi
representation for open quantum maps in
\cref{sec:quantum_maps} and for time-continuous scattering systems in
\cref{sec:time_continuous_scattering_systems}.
Specific example systems are introduced in \cref{sec:example_systems}.
We visualize their resonance states using the LR-Husimi representation in
\cref{sec:visualization_LR_Husimi} and discuss its basic properties.

\subsection{Quantum maps}
\label{sec:quantum_maps}

Quantum maps are well studied in quantum chaos and have a classical counterpart
that is a time-discrete dynamical system~\cite{HaaGnuKus2018}.
For closed quantum maps the dynamics is given by a unitary time-evolution
operator $U_\text{cl}$.
We consider quantum maps with escape for which the time-evolution operator $U$
is non-unitary.
If the system has a two-dimensional compact phase space,
e.g.\ the torus $\Gamma = [0, 1) \times [0, 1)$,
$U$ can be represented by an $N \times N$ matrix, where $N$ is the dimension of
the Hilbert space~\cite{BerBalTabVor1979, ChaShi1986}.
The size $h$ of a Planck cell in phase space is then given by $h = 1/N$.
Note that this can be straightforwardly generalized to a higher-dimensional
phase space.
The time-evolution operator of the quantum map with escape is given by
\begin{equation}
    U = U_\text{cl} \Resc,
    \label{Eq:U_definition}
\end{equation}
where $\Resc$ is a hermitian reflection operator that models the
escape~\cite{SchTwo2004, CasMasShe1999b, NonSch2008, ClaAltBaeKet2019}.

If the non-unitary time-evolution operator $U$ is also non-normal, i.e.
$U^\dagger U \neq UU^\dagger$, the left and right eigenvector of $U$ with the
same eigenvalue are not identical.
A right resonance state $| \uR_n \rangle$, $n=1, \ldots, N$, fulfills
\begin{equation}
    U |\uR_n \rangle = \lambda_n |\uR_n\rangle\, ,
    \label{Eq:right_eigenstates}
\end{equation}
with eigenvalue $\lambda_n$.
It can be expressed as
\begin{equation}
    \lambda_n = \ue^{-\frac{\gamma_n}{2}} \ue^{-\ui\theta_n},
\end{equation}
where $\gamma_n$ is the decay rate and $\theta_n$ is a phase.
The corresponding left resonance state $| \uL_n \rangle$ fulfills
\begin{equation}
    \langle \uL_n | U = \lambda_n \langle \uL_n| \, ,
    \label{Eq:left_eigenstates}
\end{equation}
with the same eigenvalue $\lambda_n$.
Alternatively, left resonance states can be considered as right resonance states
in the adjoint eigenvalue problem, $U^{\dagger} | \uL_n \rangle =
\overline{\lambda_n} | \uL_n \rangle$ with
$U^{\dagger} = \Resc U_{\text{cl}}^{-1}$ according to \cref{Eq:U_definition}
where the overbar denotes complex conjugation.

There is freedom in the choice of the normalization condition.
One possibility is
$\langle \uL_n | \uL_n \rangle = \langle \uR_n | \uR_n \rangle = 1$.
Another choice is
$\langle \uL_n | \uR_n \rangle = 1$ with
$\langle \uL_n | \uL_n \rangle = \langle \uR_n | \uR_n \rangle$.
The following equations are written such that they are valid for any
normalization.
Left and right resonance states form a biorthogonal system, i.e.
\begin{equation}
    \langle \uL_m | \uR_n \rangle = \delta_{mn} \langle \uL_n | \uR_n \rangle\,
    \label{Eq:biorthogonality}
\end{equation}
for $m, n = 1, \ldots, N$ and fulfill the completeness relation
\begin{equation}
    \sum_{n=1}^N \frac{| \uR_n \rangle \langle \uL_n |}{\langle \uL_n | \uR_n
    \rangle} = \mathbb{1},
    \label{Eq:completeness}
\end{equation}
if there is no degeneracy.
The spectral decomposition of the non-unitary time-evolution operator $U$ is
given by~\cite{ErmCarSar2009}
\begin{equation}
    U = \sum_{n=1}^N \lambda_n \frac{| \uR_n \rangle \langle \uL_n |}
    {\langle \uL_n | \uR_n \rangle} \, .
    \label{Eq:spectral_decomposition}
\end{equation}

As the structure of resonance states depends on their decay rate $\gamma_n$, we
also give some information about their distribution.
It depends on the type of escape of the system:
(i) for full escape (some columns of $U$ are zero) most decay rates are larger
than the natural decay rate $\gnat$ from classical
dynamics~\cite{GasRic1989b, She2008, NonZwo2009, Nov2012,
BarWeiPotStoKuhZwo2013, Vac2024},
(ii) for partial escape most decay rates are in the interval $[\gnat, \ginv]$
with $\ginv$ being the inverse natural decay rate of the time-inverted classical
dynamics~\cite{NonSch2008, GutOsi2015, AltPorTel2015, ClaAltBaeKet2019}.
In the semiclassical limit they are conjectured to accumulate around the
typical decay rate $\gtyp$~\cite{NonSch2008}.

For the comparison of quantum resonance states with the underlying classical
dynamics one needs a phase-space representation of resonance states.
Here, we use the Husimi representation~\cite{Hus1940, Lee1995} which can be
interpreted as a probability density on phase space.
For a right resonance state $|\uR_n \rangle$ it is given by
\begin{equation}
    \HusR_n(\vecx) =
    \frac{| \langle \alpha(\vecx) | \uR_n \rangle |^2}
    { \int_\Gamma \ud \vecx' \, | \langle \alpha(\vecx') | \uR_n \rangle |^2 }\, ,
    \label{Eq:definition_husimi_right}
\end{equation}
where $|\alpha(\vecx) \rangle$ is a coherent state centered at a phase-space
point $\vecx = (q, p) \in \Gamma$ with normalization
$\langle \alpha(\vecx) | \alpha(\vecx) \rangle~=~1$.
In \cref{Eq:definition_husimi_right} we use a normalization of the Husimi
representation such that $\int_{\Gamma} \ud \vecx\, \HusR_n(\vecx) = 1$.
Equivalently, one can define the Husimi representation of the left resonance
state $\langle \uL_n |$ as
\begin{equation}
    \HusL_n(\vecx) =
    \frac{| \langle \uL_n | \alpha(\vecx) \rangle |^2}
    { \int_\Gamma \ud \vecx' \, | \langle \uL_n | \alpha(\vecx') \rangle |^2 }\, .
    \label{Eq:definition_husimi_left}
\end{equation}

A combined representation from left and right resonance states uses the
complex-valued \textit{left-right Husimi amplitude} (LR-Husimi amplitude)
\begin{equation}
    \hLR_n(\vecx) = \frac{\langle \alpha(\vecx) | \uR_n \rangle
    \langle \uL_n | \alpha(\vecx) \rangle}{\langle \uL_n | \uR_n \rangle} \, .
    \label{Eq:LR_husimi_amplitude}
\end{equation}
From this the \textit{left-right Husimi representation} (LR-Husimi
representation) is obtained by
\begin{equation}
    \HusLR_n(\vecx) = \frac{|\hLR_n(\vecx)|}{\int_\Gamma \ud \vecx' \,
    |\hLR_n(\vecx')|}\, ,
    \label{Eq:LR_husimi_definition}
\end{equation}
which was introduced in Ref.~\cite{ErmCarSar2009}.
We define the LR-Husimi representation to be normalized,
$\int_\Gamma \ud \vecx\, \HusLR_n(\vecx) = 1$,
such that it can be interpreted as a probability density on phase space.
The LR-Husimi representation can be expressed directly in terms of the
Husimi representations of the left and right resonance
states~\cite{ErmCarSar2009},
\cref{Eq:definition_husimi_right,Eq:definition_husimi_left},
\begin{equation}
    \HusLR_n(\vecx) = \frac{\sqrt{\HusL_n(\vecx) \HusR_n(\vecx)}}
        {\int_\Gamma \ud \vecx' \,  \sqrt{\HusL_n(\vecx') \HusR_n(\vecx')}}\, ,
    \label{Eq:LR_husimi}
\end{equation}
which will be used in the rest of this paper.

For quantum maps with time-reversal symmetry, specifically $U_\text{cl} =
U_\text{cl}^{T}$ in position space, the left resonance states are given by
$L_{n,q} = \sum_{q'} (U_\text{cl})_{q,q'} \overline{R_{n,q'}}$ (this differs
from Ref.~\cite{KeaNovSch2008}, as we use an interchanged order of $U_\text{cl}$
and $\Resc$ in \cref{Eq:U_definition}).
One can show that therefore the left Husimi representation is approximately
related to the right Husimi representation by
$\HusL_n(q, p) \approx \Resc(q, p) \HusR_n(q, 1-p)$, where $\Resc(q, p)$ is the
classical reflection at phase space point $(q, p)$.
This approximation becomes better in the semiclassical limit, where the
reflection operator $\Resc$ applied to a coherent state is well approximated by
the classical reflection function evaluated at the center of the coherent state.
For the LR-Husimi representation,~\cref{Eq:LR_husimi}, this implies an
approximate mirror symmetry in $p$, $\HusLR(q,p) \approx \HusLR(q,1-p)$, if
additionally the reflection is symmetric, $\Resc(q,p) = \Resc(q, 1-p)$.

\subsection{Time-continuous scattering systems}
\label{sec:time_continuous_scattering_systems}

Another class of systems which are especially of interest for experimental
realizations are time-continuous scattering systems, in particular
two-dimensional billiards with escape like dielectric cavities~\cite{CaoWie2015}
or the three-disk scattering system~\cite{GasRic1989c, PotWeiBarKuhStoZwo2012,
BarWeiPotStoKuhZwo2013, WeiBarKuhPolSch2014, CviArtMaiTan2020}.
In position space, resonance states are solutions of the Helmholtz equation
fulfilling outgoing boundary conditions,
\begin{equation}
    \Delta \psiR(\vecr) = - n^2(\vecr) \, k^2 \, \psiR(\vecr)\, ,
    \label{Eq:Helmholtz_right}
\end{equation}
for right resonance states $\psiR(\vecr)$, where $\Delta$ is the Laplace
operator in two dimensions, $n(\vecr)$ is the local refractive index, and $k$ is
the complex wavenumber.
The imaginary part of $k$ is proportional to the dimensionless decay rate
$\gamma$~\cite{CaoWie2015, SchKet2023}.
Left resonance states $\psiL(\vecr)$ fulfill an incoming boundary condition
and
\begin{equation}
    \Delta \overline{\psiL(\vecr)} = - n^2(\vecr) \, k^2 \,
    \overline{\psiL(\vecr)} \,.
    \label{Eq:Helmholtz_left}
\end{equation}

The LR-Husimi representation for time-continuous scattering systems is obtained
by using the Husimi representation for right and left resonance states, see
\cref{Eq:LR_husimi}.
For dielectric cavities the Husimi representation for right and left resonance
states is defined in Ref.~\cite{HenSchSch2003}.
Specifically, for right and left resonance states we use the incoming Husimi
representation $\Hus_n^\text{inc}$ inside the cavity.
Alternatively, one could use for both states the emanating Husimi
representation, giving the same LR-Husimi representation in the semiclassical
limit.
In this context the terms incoming and emanating refer to the definition of the
boundary Husimi representation inside the cavity in Ref.~\cite{HenSchSch2003}
and are not related to the incoming and outgoing boundary conditions of left and
right resonance states.
For the three-disk scattering system the Husimi representation is defined in
Ref.~\cite{WeiBarKuhPolSch2014}, where also the LR-Husimi representation is
shown.

In analogy to the discussion of time-reversal symmetry for quantum maps, for the considered time-continuous scattering systems one has
$\psiL_n(\vecr) = \overline{\psiR_n(\vecr)}$~\cite{SteWal1972,
WeiBarKuhPolSch2014}.
Therefore, the left Husimi representation is approximately
$\HusL_n(q,p) \approx \Resc(q,p) \HusR_n(q,-p)$.
For the LR-Husimi representation,~\cref{Eq:LR_husimi}, this implies an
approximate mirror symmetry in $p$, $\HusLR(q,p) \approx \HusLR(q,-p)$ using
that the reflection $\Resc(q,p)$ is symmetric in $p$ for billiard dynamics.

\subsection{Example systems}
\label{sec:example_systems}

We use four exemplary chaotic scattering systems throughout this
paper~\cite{KetLorSch2025}.
These examples cover the cases of partial and full escape for a 2D billiard and
a map in each case and have a fully chaotic phase space:
\begin{itemize}
    \item[(a)] Dielectric cavity (partial escape) with lima\c{c}on shape,
        deformation $\epsilon = 0.6$, where it is practically fully chaotic,
        radius $\Rcav$, refractive index $n_\mathrm{r}=3.3$, and reflection law
        for a TM polarized mode~\cite{KetClaFriBae2022}.
        The dimensionless decay rate $\gamma$ is related to the wavenumber $k$
        by $\gamma = -2 \, \Imag \, kR_\text{cav}$.
    \item[(b)] Three-disk scattering system (full escape) with disks of radius
        $a$ and center to center distance $2.1 a$~\cite{SchKet2023}.
        The dimensionless decay rate is $\gamma = -2 \, \Imag \, k a$.
    \item[(c)] Standard map at kicking strength $K = 10$, where it is practically
        fully chaotic, with partial reflectivity $0.2$ in the interval
        $q \in [0.3, 0.6]$~\cite{ClaAltBaeKet2019}.
    \item[(d)] Like (c) but with reflectivity 0 (full
        escape)~\cite{ClaAltBaeKet2019}.
\end{itemize}
All example systems have time-reversal symmetry (in the closed system) and the reflection has a mirror symmetry in $p$.
For more details on the specific systems used here, see the supplemental
material of Ref.~\cite{KetLorSch2025}.

In all figures, for the most semiclassical resonance states we use the following
ranges of wavenumbers:
(a) for the dielectric cavity $\Real\, k\Rcav \in [5\,000, 5\,025]$,
(b) for the three-disk scattering system $\Real\, ka \in [100\,000, 100\,200]$
(except for $\gamma = 1.8$ in
Figs.~\ref{fig:fluctuations}, \ref{fig:husimi_ray_wave_comparison}, and
\ref{fig:husimi_ray_wave_comparison_appendix},
where
$\Real\, ka \in [50\,000, 50\,500]$ is used), and
(c, d) for the standard map with partial and full escape all even
$N \in [19\,996, 20\,004]$.

We expect that our results hold for higher-dimensional billiards and maps with
fully chaotic phase spaces just as well, since all our arguments are independent
of the number of dimensions of phase space.
However, those systems are harder to visualize and numerically more challenging,
in particular they would not allow us to go as far into the semiclassical limit.
Therefore, we focus on the lowest dimensional billiards and maps with chaotic
dynamics.

\subsection{Visualization of LR-Husimi representation}
\label{sec:visualization_LR_Husimi}

\begin{figure*}
    \begin{center}
        \includegraphics[width=1\linewidth]{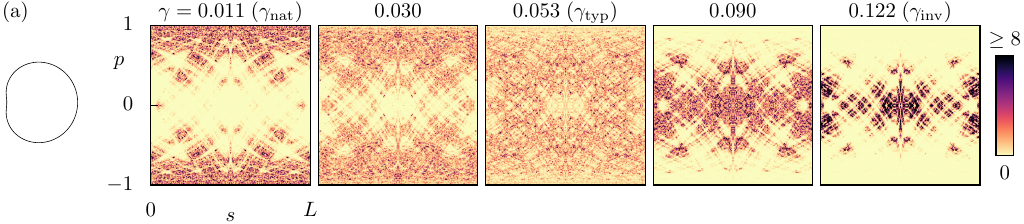}
        \vspace{0.1cm}

        \includegraphics[width=1\linewidth]{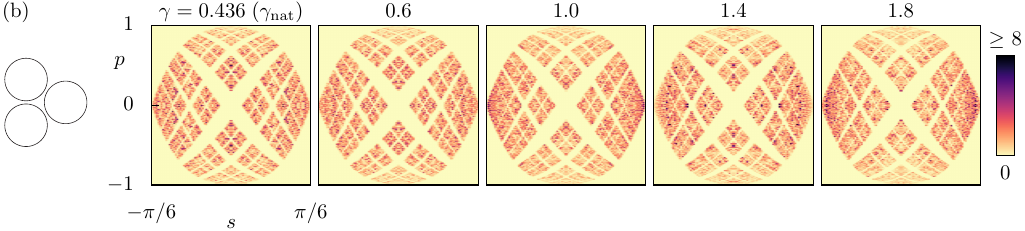}
        \vspace{0.1cm}

        \includegraphics[width=1\linewidth]{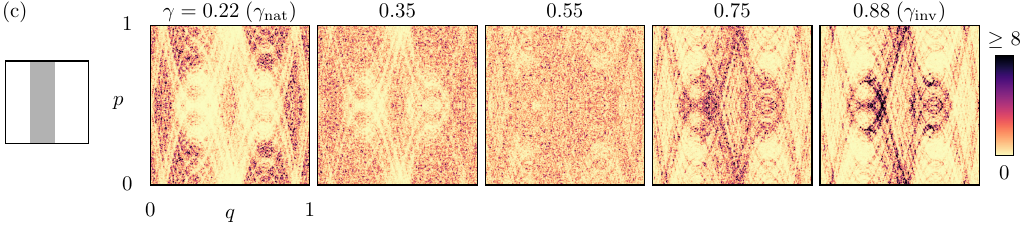}

        \vspace{0.3cm}
        \includegraphics[width=1\linewidth]{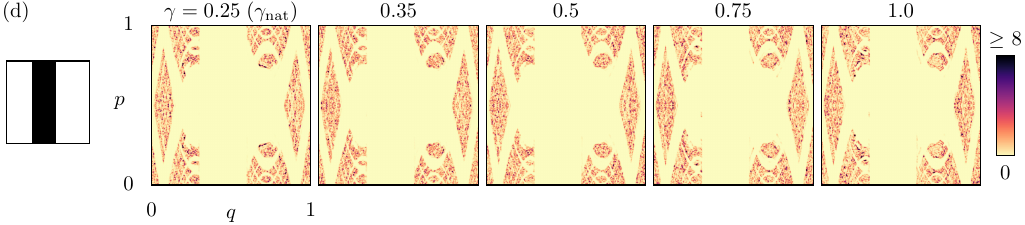}
    \end{center}
    \caption{LR-Husimi representation $\HusLR$ of individual resonance states
    closest to the specified decay rate $\gamma$ for
    (a) a dielectric cavity,
    (b) the three-disk scattering system,
    (c) the standard map with partial escape,
    and (d) the standard map with full escape.
    System parameters are given in \cref{sec:example_systems}.
    In this and all following figures the average intensity is scaled to one (in
    (a) and (c) on the full phase space and in (b) and (d)  on the chaotic saddle).
    Intensities greater than the maximal value of the colorbar are shown with darkest color.
    }
    \label{fig:husimi_individual}
\end{figure*}

\cref{fig:husimi_individual} shows the LR-Husimi representation $\HusLR$ for resonance
states of all example systems for various decay rates.
Due to the above discussed symmetry in all example systems, the LR-Husimi representation is approximately mirror symmetric in $p$.

For systems with partial escape, \cref{fig:husimi_individual}(a, c), the LR-Husimi
representation is supported by the entire phase space.
In contrast, for systems with full escape, \cref{fig:husimi_individual}(b, d),
it is localized on the chaotic saddle.
This can be understood as follows:
The Husimi representation of left (right) resonance states is supported by the
forward (backward) trapped set~\cite{CasMasShe1999b, KeaNovPraSie2006}.
The LR-Husimi representation is a product of both Husimi representations,
\cref{Eq:LR_husimi}, and hence is supported by the intersection of both sets,
which is the chaotic saddle (sometimes termed repeller).

Furthermore, for systems with full escape, \cref{fig:husimi_individual}(b, d),
the LR-Husimi representation is almost independent of the decay rate $\gamma$
compared to the strong dependence on $\gamma$ for partial escape,
\cref{fig:husimi_individual}(a, c).
More precisely, for all decay rates it is constant on the chaotic saddle up to
fluctuations (with a $\sqrt{1 - p^2}$ dependence for the three-disk scattering
system due to the chosen definition of the boundary Husimi
representation~\cite{BaeFueSch2004}).

\clearpage

\section{Quantum invariance}
\label{sec:invariance}

\begin{figure}
	\begin{center}
		\includegraphics{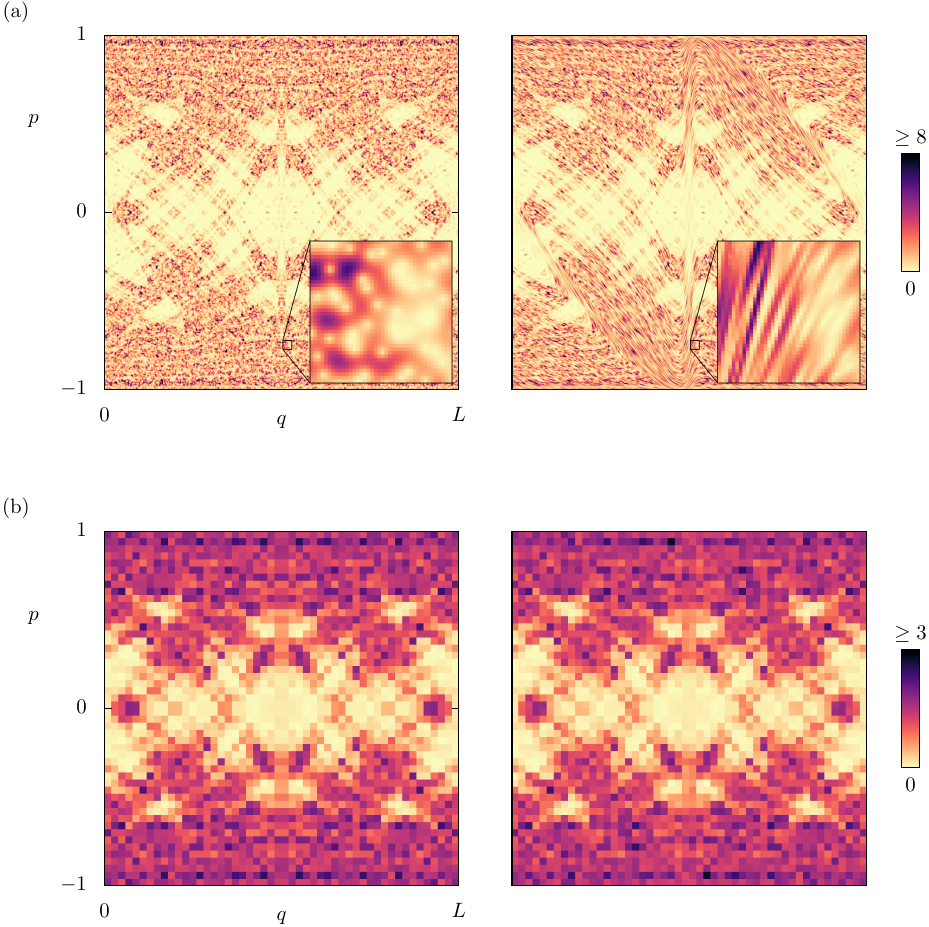}
	\end{center}
	\caption{
    Visual comparison of the LR-Husimi representation $\HusLR$
    according to left-hand side (left) and right-hand side (right)
    of \cref{Eq:LR_husimi_invariance}.
    The integration is done over regions $A$ from
    (a) a fine $1600 \times 1600$ grid showing deviations from invariance
    due to stretching along the unstable direction,
    see zoom by factor 16,
    and
    (b) a coarse $50 \times 50$ grid showing less deviations from invariance (even though we choose a more sensitive range for the colorbar).
    The presented individual resonance state is closest to $\gamma=0.03$
    for a dielectric cavity, see \cref{fig:husimi_individual}(a, second panel).
	}
	\label{fig:husimi_invariance}
\end{figure}
\begin{figure}
	\begin{center}
		\includegraphics{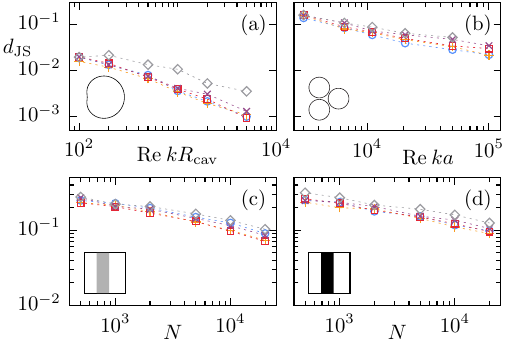}
	\end{center}
	\caption{Semiclassical convergence of left-hand side to
        right-hand side of \cref{Eq:LR_husimi_invariance}
        for a fixed size of regions $A$ of a $50 \times 50$ grid
        demonstrating quantum invariance.
        Shown is the Jensen-Shannon divergence $d_{\text{JS}}$
        evaluated on this grid approaching zero in the semiclassical
        limit.
        (a-d) Four example systems with resonance states near
        five decay rates each, as in \cref{fig:husimi_individual}.
        Symbols as in \cref{fig:JS_n_grid}.
	}
	\label{fig:JS_husimi_invariance}
\end{figure}

We demonstrate that resonance states in the LR-Husimi
representation are invariant under the closed dynamics
in the semiclassical limit.
We call this \textit{quantum invariance}
in analogy to quantum ergodicity,
as explained below.
First we present visual and quantitative numerical evidence,
later we give analytical support.

Invariance of some phase-space distribution $\rho(\vecx)$ under a map $T$
is defined by
\begin{equation}
    \int_A \ud \vecx \, \rho(\vecx)
    =
    \int_{T^{-1}(A)} \ud \vecx \, \rho(\vecx)
    \, ,
    \label{Eq:invariance}
\end{equation}
for any phase-space region $A$,
where $T^{-1}(A)$ is the preimage of $A$.
In the present context the classical map $T$ corresponds to the unitary
time-evolution operator $U_{\text{cl}}$ of the closed quantum map.
Note that $T$ also can be the boundary map of a billiard system.
The time it takes from one to the next reflection in the billiard
has no relevance for invariance.

When specifying such an invariance for the LR-Husimi representation $\HusLR$,
\begin{equation}
    \int_A \ud \vecx \, \HusLR(\vecx)
    \approx
    \int_{T^{-1}(A)} \ud \vecx \, \HusLR(\vecx)
    \, ,
    \label{Eq:LR_husimi_invariance}
\end{equation}
it is at best approximately true.
This is obvious for regions $A$ smaller than or of the same size as a coherent state,
i.e.\ the scale $h$ of a Planck cell:
The Husimi representation is
smooth on this scale and it has local maxima and zeros with the same
elliptic shape as the coherent states used in the representation.
Under the hyperbolic dynamics this shape becomes stretched
and leads to deviations from invariance in \cref{Eq:LR_husimi_invariance}.
This is visualized
for a resonance state of the dielectric cavity in
\cref{fig:husimi_invariance}, comparing the left-hand side
and right-hand side of \cref{Eq:LR_husimi_invariance}
shown on the left and the right, respectively.
In \cref{fig:husimi_invariance}(a)
we evaluate \cref{Eq:LR_husimi_invariance}
for regions $A$ of a fine $1600 \times 1600$ grid
in phase space, i.e.\ $A \ll h$.
On the right one observes a stretching
along the unstable direction, best seen in the zoom region.
In this case one indeed finds deviations from invariance.
In contrast, in \cref{fig:husimi_invariance}(b)
we use regions $A$ of a coarse $50 \times 50$ grid.
For these larger regions $A \gg h$ the stretched shape on the right
is no longer resolved.
The left-hand side and right-hand side of
\cref{Eq:LR_husimi_invariance}
appear to be much closer.

This is analyzed quantitatively
for a fixed size of the regions $A$
when going to the semiclassical limit, $h \rightarrow 0$,
choosing resonance states near a fixed decay rate $\gamma$,
see \cref{fig:JS_husimi_invariance}.
Specifically, we use regions $A$ from a $50 \times 50$ grid
and quantify the difference between the
left-hand side and right-hand side of \cref{Eq:LR_husimi_invariance},
as visualized in \cref{fig:husimi_invariance}(b),
using the Jensen-Shannon divergence~\cite{Lin1991}.
In the semiclassical limit,
i.e.\ larger wavenumber $k$ or larger matrix size $N$,
it decays to zero for all example systems and decay rates.

We thus define \textit{quantum invariance}
as a property of resonance states
in their LR-Husimi representation $\HusLR$:
In the semiclassical limit $h \rightarrow 0$
and for a phase-space region $A$ with a fixed size,
there is invariance under the closed dynamics,
i.e.\ the left-hand side and right-hand side of
\cref{Eq:LR_husimi_invariance} approach each other.
We use the term quantum invariance
in analogy to the term quantum ergodicity,
where a property from dynamical systems is attributed to
a quantum system under certain qualifications
in the semiclassical limit.

We mention as an aside that we attribute the different magnitude of the
Jensen-Shannon divergence in \cref{fig:JS_husimi_invariance}
to the different magnitude
of the Lyapunov exponent in the various systems.
In particular, the Lyapunov exponent is the smallest for the lima\c{c}on shaped
dielectric cavity, leading to the least stretching and thus to the
least deviations from invariance, i.e.\ the
smallest value of the Jensen-Shannon divergence.

In the remainder of this section we demonstrate analytically
that the LR-Husimi representation is approximately
invariant under the closed dynamics,
supporting quantum invariance.
We discuss the case of a quantum map from \cref{sec:quantum_maps}
with a time-evolution operator
according to \cref{Eq:U_definition}, $U = U_\text{cl} \Resc$, where
$U_\text{cl}$ is the unitary time-evolution operator of the closed map.
In the numerator of the definition of the LR-Husimi amplitude, \cref{Eq:LR_husimi_amplitude},
(i) we first insert the identity $I = U_\text{cl} U_\text{cl}^{-1}$ twice, leading to
\begin{align}
    \langle \alpha(\vecx) | \uR_n \rangle \langle \uL_n | \alpha(\vecx) \rangle
    &\stackrel{(\text{i})}{=} \nonumber
    \langle \alpha(\vecx) | U_\text{cl}
    U_\text{cl}^{-1} |\uR_n \rangle \langle \uL_n | U_\text{cl}
    U_\text{cl}^{-1} | \alpha(\vecx) \rangle
    \\
    &\stackrel{(\text{ii})}{\approx}  \nonumber
    \langle \alpha(T^{-1}(\vecx)) |
    U_\text{cl}^{-1} |\uR_n \rangle \langle \uL_n | U_\text{cl}
    | \alpha(T^{-1}(\vecx)) \rangle
    \\
    &\stackrel{(\text{iii})}{=}  \nonumber
    \langle \alpha(T^{-1}(\vecx)) |
    \Resc   |\uR_n \rangle \langle \uL_n | \Resc^{-1}
    | \alpha(T^{-1}(\vecx)) \rangle
    \\
    &\stackrel{(\text{iv})}{\approx}
    \langle \alpha(T^{-1}(\vecx)) |\uR_n \rangle
    \langle \uL_n | \alpha(T^{-1}(\vecx)) \rangle
    \, .
    \label{Eq:LR_husimi_amplitude_invariant_long}
\end{align}
In step (ii) we use the approximate identity
\begin{equation}
    U_\text{cl}^{-1} | \alpha(\vecx) \rangle
    \approx
    | \alpha(T^{-1}(\vecx)) \rangle
    \, ,
\end{equation}
that the backward time-evolved coherent state is (up to squeezing)
the coherent state at the backward time-evolved phase-space position.
In step (iii) we use the identity
\begin{equation}
    U_\text{cl}^{-1} |\uR_n \rangle \langle \uL_n | U_\text{cl}
    =
    \Resc   |\uR_n \rangle \langle \uL_n | \Resc^{-1}
    \, ,
\end{equation}
which is based on the definition of right and left eigenstates
(product of applying $U_\text{cl}^{-1}$ from the left to \cref{Eq:right_eigenstates}
and $\Resc^{-1}$ from the right to \cref{Eq:left_eigenstates};
for systems with full escape, where $\Resc^{-1}$ is not defined,
one has to suitably restrict the argument to the chaotic saddle).
In the last step (iv) of \cref{Eq:LR_husimi_amplitude_invariant_long}
we use that the reflection function is approximately constant
on the scale of the coherent state. Then for $\vecy=T^{-1}(\vecx)$
the term $\Resc^{-1} | \alpha(\vecy) \rangle$
multiplies the coherent state with a factor that is the inverse of the factor from
$\langle \alpha(\vecy)) | \Resc$ such that they cancel.

From \cref{Eq:LR_husimi_amplitude_invariant_long} follows
the approximate pointwise invariance of the smooth
LR-Husimi amplitude, \cref{Eq:LR_husimi_amplitude},
\begin{align}
    &\hLR_n(\vecx) \approx  \hLR_n(T^{-1}(\vecx))
    \, .
    \label{Eq:LR_husimi_amplitude_invariant}
\end{align}
By integrating over some phase-space region $A$,
we find the approximate invariance,
\begin{align}
    \int_A \ud \vecx \, \hLR_n(\vecx)
    \approx  \nonumber
    \int_A \ud \vecx \, \hLR_n(T^{-1}(\vecx))
    =
    \int_{T^{-1}(A)} \ud \vecy \, \hLR_n(\vecy)
    \, .
    \label{Eq:LR_husimi_region_invariant}
\end{align}
Similarly, one finds for the LR-Husimi representation,
using $\HusLR_n(\vecx) \propto |\hLR_n(\vecx)|$
according to \cref{Eq:LR_husimi},
an approximate invariance corresponding to \cref{Eq:LR_husimi_invariance}.
In order to demonstrate quantum invariance one would have to show that
in the semiclassical limit,
when the size of the coherent state is arbitrarily small
compared to the region $A$,
the relative error of the used approximations goes to zero
when integrating over $A$.

\section{Factorization}
\label{sec:factorization}

In this section we show that the LR-Husimi representation of resonance states
can be factorized into a product of a density from classical dynamics and
universal fluctuations.
To this end we first review the factorization for
right and left resonance states in~\cref{sec:factorization_resonance_states}.
Based on this we derive the factorization for the LR-Husimi
in~\cref{sec:factorization_LR_husimi} and analyze both factors in detail
in Secs.~\ref{sec:distribution_LR_fluctuations}
and~\ref{sec:construction_LR_measure}.

\subsection{Factorization conjecture of resonance states}
\label{sec:factorization_resonance_states}

According to a recently proposed factorization
conjecture~\cite{ClaKunBaeKet2021, KetClaFriBae2022, SchKet2023}, right
resonance states in chaotic scattering are a product of (i) a conditionally
invariant measure from classical dynamics which by smoothing on the scale of a
Planck cell gives a smooth density $\rhoR$
depending on the decay rate $\gamma$ and
(ii) universal exponentially
distributed fluctuations with mean one.
This factorization holds for the intensity $|\langle \varphi | \uR \rangle|^2$
of right resonance states and is independent of the chosen (possibly
over-complete) basis $|\varphi\rangle$~\cite{ClaKunBaeKet2021}.
For example, using coherent states $|\varphi\rangle = |\alpha(\vecx)\rangle$,
the Husimi representation of a right resonance state with decay rate $\gamma$ is
given by the product
\begin{equation}
    \HusR(\vecx) = \rhoR(\vecx) \cdot \etaR(\vecx)\, .
    \label{Eq:factorization_right}
\end{equation}
The fluctuations $\etaR$, statistically independent from $\rhoR$, follow an
exponential distribution $P(\etaR)= \exp(-\etaR)\,$.

Since left resonance states are right resonance states of a modified system
they factorize as well,
\begin{equation}
    \HusL(\vecx) = \rhoL(\vecx) \cdot \etaL(\vecx)\, ,
    \label{Eq:factorization_left}
\end{equation}
with fluctuations $\etaL$, statistically independent from $\rhoL$, with the same
exponential distribution $P(\etaL)= \exp(-\etaL)\,$.

\subsection{Factorization of LR-Husimi representation}
\label{sec:factorization_LR_husimi}
In the following, we derive such a factorization also for the LR-Husimi
representation based on the factorization conjecture for left and right resonance states.
To this end we start with \cref{Eq:LR_husimi} and insert
\cref{Eq:factorization_right,Eq:factorization_left} to find
\begin{align}
    \HusLR(\vecx) = \frac{\sqrt{\rhoL(\vecx) \rhoR(\vecx)}\ \sqrt{\etaL(\vecx)
    \etaR(\vecx)}}{\int_\Gamma \ud \vecx' \sqrt{\rhoL(\vecx') \rhoR(\vecx')}\
    \sqrt{\etaL(\vecx') \etaR(\vecx')}}
\end{align}
We now use from the factorization conjecture above
that the right fluctuations $\etaR(\vecx)$ are
statistically independent of their classical density $\rhoR(\vecx)$.
The same is true for the left fluctuations $\etaL(\vecx)$ and their classical
density $\rhoL(\vecx)$.
Hence, the factor $\sqrt{\etaL(\vecx') \etaR(\vecx')}$ in the
denominator can be replaced by its expectation value.
This leads to the factorization for the LR-Husimi representation
\begin{equation}
    \HusLR(\vecx) = \rhoLR(\vecx) \cdot \etaLR(\vecx)\, ,
    \label{Eq:factorization_LR_husimi}
\end{equation}
where the first factor is a smooth density,
\begin{equation}
    \rhoLR(\vecx) = \frac{\sqrt{\rhoL(\vecx) \rhoR(\vecx)}}{\int_\Gamma \ud
    \vecx' \sqrt{\rhoL(\vecx') \rhoR(\vecx')}}
    \, .
    \label{Eq:LR_density}
\end{equation}
It is given by the square root of the product of the smooth densities
$\rhoL(\vecx)$ and $\rhoR(\vecx)$ that describe the left and right resonance
states, respectively.
Below in \cref{sec:construction_LR_measure}, we will show that there is a
measure $\muLR$ constructed from classical dynamics which by smoothing
on the scale of a Planck cell gives $\rhoLR(\vecx)$.

We call the second factor in \cref{Eq:factorization_LR_husimi} the
LR-fluctuations, defined by
\begin{equation}
    \etaLR = \frac{\sqrt{\etaL \etaR}}{\text{E} \left[ \sqrt{\etaL \etaR}
    \right]}\, .
    \label{Eq:LR_fluctuations}
\end{equation}
It is given by the square root of the product of two exponentially distributed
random variables $\etaL$ and $\etaR$ with mean one.
From their definition it follows directly that their expectation value is one,
$\text{E} \left[ \etaLR \right] = 1$.
Averaging \cref{Eq:factorization_LR_husimi} over resonance states with similar
decay rates $\gamma$ thus relates the smooth density $\rhoLR(\vecx)$ to the
averaged LR-Husimi representation,
\begin{equation}
    \langle \HusLR(\vecx) \rangle_\gamma = \rhoLR(\vecx)\, .
    \label{Eq:averaged_LR_husimi}
\end{equation}

\begin{figure*}
    \begin{center}
        \includegraphics{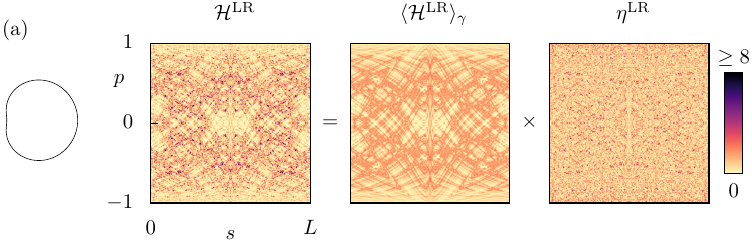}
        \includegraphics{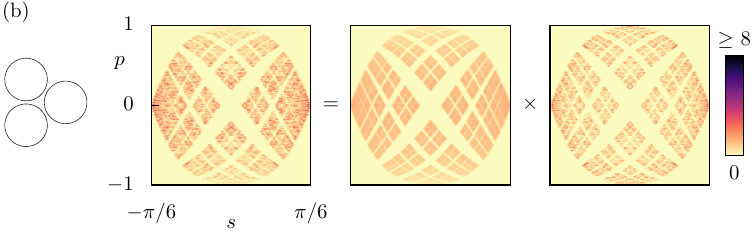}
        \includegraphics{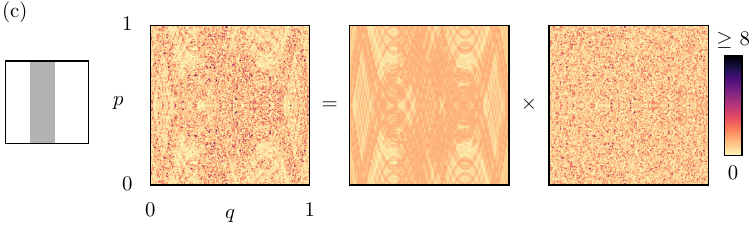}
        \includegraphics{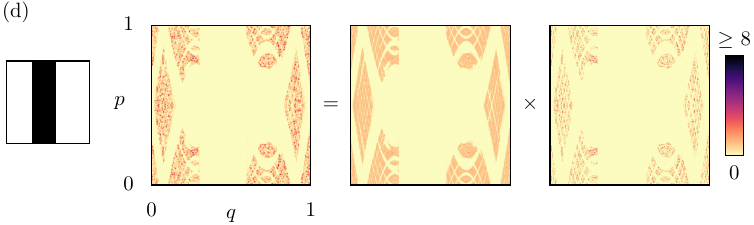}
    \end{center}
    \caption{Factorization of LR-Husimi representation $\HusLR$ (left) according
    to \cref{Eq:LR_Husimi_decomposition} into the average
    $\langle \HusLR \rangle_\gamma$  of resonance states close to decay rate
    $\gamma$ (middle) times LR-fluctuations $\etaLR$,
    \cref{Eq:LR_fluctuations_numerical}, (right) for
    (a) dielectric cavity at $\gamma = 0.053$,
    (b) three-disk scattering system at $\gamma = 1.0$,
    (c) standard map with partial escape at $\gamma = 0.055$, and
    (d) standard map with full escape at $\gamma = 0.5$.
    }
    \label{fig:husimi_factorization}
\end{figure*}

In \cref{fig:husimi_factorization} we illustrate the factorization of the
LR-Husimi representation for a resonance state for all example system, with the
numerical approach explained in \cref{sec:numerical_verification}.
The first factor (middle column) will be compared to a
measure $\muLR$ in \cref{sec:comparison_LR_measure_Husimi}.
The LR-fluctuations (right column) appear to be the same in the entire phase
space for partial escape, \cref{fig:husimi_factorization}(a, c), and
on the chaotic saddle for full escape, \cref{fig:husimi_factorization}(b, d).
Given that the systems are very different, this suggests that the
LR-fluctuations are universal.
We derive their distribution in \cref{sec:distribution_LR_fluctuations} and
numerically support the universality in \cref{sec:numerical_fluctuations}.

\subsection{Distribution of LR-fluctuations}
\label{sec:distribution_LR_fluctuations}
We now derive the distribution of the LR-fluctuations,
\cref{Eq:LR_fluctuations}.
They are defined by the square root of the product of two exponentially
distributed random variables $\etaL$ and $\etaR$ with mean one
and are normalized such that $\text{E} \left[ \etaLR \right] = 1$.

In the following we make the assumption of statistical independence of the left
and right fluctuations, $\etaL$ and $\etaR$.
Note that for systems with time-reversal symmetry (in the closed system) this is
not applicable in position space, where left and right resonance states are
complex conjugates of each other and therefore their fluctuations are maximally
correlated.

First, we evaluate the denominator in~\cref{Eq:LR_fluctuations},
\begin{equation}
    \text{E} \left[ \sqrt{\etaL \etaR} \right]
    =
    \left( \int_0^{\infty} \ud \eta \, P(\eta) \, \sqrt{\eta} \right)^2
    = \frac{\pi}{4}\, ,
\end{equation}
where $P(\eta) = \exp(-\eta)$.
The probability distribution $\PLR(\etaLR)$ is then given by
\begin{align}
            \PLR(\etaLR) =
    \int_{0}^\infty \ud \etaL
    \int_{0}^\infty \ud \etaR
    \, P(\etaL) \, P(\etaR)
    \; \; \delta\left(\etaLR - \frac{4}{\pi}
    \sqrt{\etaL \etaR} \right)\, .
\end{align}
For one integration the $\delta$-function can be exploited,
while the other gives an
integral representation of $K_0$,
the $0$-th order modified Bessel function of the second
kind~\cite{GraRyz2014},
leading to the universal distribution of LR-fluctuations,
\begin{equation}
    \PLR(\etaLR)
    =
    \frac{\pi^2}{4} \, \etaLR \, K_0 \left( \frac{\pi}{2} \etaLR \right)
    \, .
    \label{Eq:LR_fluctuations_distribution}
\end{equation}

\begin{figure}[b]
    \begin{center}
        \includegraphics{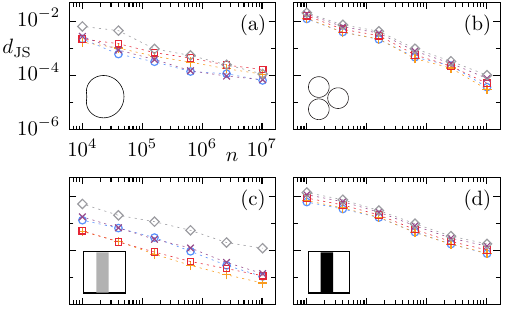}
    \end{center}
    \caption{
        Convergence of coarse-grained LR-measure $\muLR$ for
        increasingly fine partitions with $n$ cells.
        Shown is the Jensen-Shannon divergence $d_{\text{JS}}$ (evaluated on a
        $50 \times 50$ grid) between measures from partitions $n$ and $n/4$
        approaching zero.
        (a) Dielectric cavity for
        $\gamma \in \{ 0.011 \, (\gnat), 0.030, 0.053 \, (\gtyp), 0.090, 0.122 \, (\ginv)
        \}$.
        (b) Three-disk scattering system for
        $\gamma \in \{0.436 \, (\gnat), 0.6, 1.0, 1.4, 1.8 \}$.
        (c) Standard map with partial escape for
        $\gamma \in \{0.22 \, (\gnat), 0.35, 0.55, 0.75, 0.88 \, (\ginv) \}$.
        (d) Standard map with full escape for
        $\gamma \in \{0.25 \, (\gnat), 0.35, 0.5, 0.75, 1.0 \}$.
        Symbols {\large $\circ$}, $+$, {\scriptsize $\square$}, $\times$, and
        {\large $\diamond$} are used for increasing $\gamma$.
    }
    \label{fig:JS_n_grid}
\end{figure}

\subsection{Construction of LR-measure}
\label{sec:construction_LR_measure}

Next, we propose the construction of a measure $\muLR$,
which we call LR-measure.
This multifractal measure, after smoothing on the scale of a Planck cell, gives
the smooth density $\rhoLR(\vecx)$,~\cref{Eq:LR_density}.
In \cref{sec:comparison_LR_measure_Husimi} the LR-measure is compared to the
first factor of the factorization in \cref{fig:husimi_factorization}.

In Ref.~\cite{KetLorSch2025} we proposed the construction of a conditionally
invariant measure $\muR$ that describes the average of right resonance states
with decay rate $\gamma$ in the semiclassical limit.
Its construction generalizes Ulam's method~\cite{Ula1960} for finding the
leading eigenvector of a matrix approximation of the Perron-Frobenius
operator~\cite{CviArtMaiTan2020}.
This is done by allowing more general matrix approximations of the
operator~\cite{KetLorSch2025}.
From the infinite set of matrix approximations one has to select the one that
fulfills an optimization criterion~\cite{KetLorSch2025}.
The conditionally invariant measure $\muR$ is then given by the leading
eigenvector of the selected matrix approximation.
Numerically, for a partition of phase space into $n$ disjoint cells
$\{A_1, \dots, A_n\}$ one determines the coarse-grained conditionally invariant
measure $\mu_{\gamma, i}^\mathrm{R} = \muR(A_i)$.
In the limit of increasingly finer partitions of phase space,
$n \rightarrow \infty$, the right measures $\muR$ were found to
converge~{\cite{KetLorSch2025}}.

The proposed method can also be used to find a conditionally invariant
measure $\muL$ that describes left resonance states.
Since left resonance states are right resonance states of $U^{\dagger}$, see
discussion after \cref{Eq:left_eigenstates}, one uses the time-reversed closed
dynamics and applies the reflection afterwards.
For systems with time-reversal symmetry (in the closed system), the left measure
$\muL$ can be directly obtained from the right measure in phase space by
mirroring in $p$ and subsequently applying the reflection.

We define the coarse-grained LR-measure
$\mu_{\gamma, i}^\mathrm{LR} = \muLR(A_i)$ for each cell $A_i$ by box-wise
multiplication of the left and right measures and normalization,
\begin{equation}
    \mu_{\gamma, i}^\mathrm{LR}
    = \frac{\sqrt{\mu_{\gamma, i}^\mathrm{L} \ \mu_{\gamma, i}^\mathrm{R}}}
            {\sum_{i'=1}^n \sqrt{\mu_{\gamma, i'}^\mathrm{L} \
            \mu_{\gamma, i'}^\mathrm{R}}}\, .
            \label{Eq:LR_measure}
\end{equation}
The square root is taken in accordance to \cref{Eq:LR_density}.
Note that we use boxes of equal size, otherwise one would have to suitably adopt
the normalization.

In the limit of increasingly finer partitions of phase space,
$n \rightarrow \infty$, we find convergence for the LR-measure $\muLR$,
see~\cref{fig:JS_n_grid}.
We numerically demonstrate the convergence by comparing the coarse-grained
measures for $n$ and for $n/4$ cells using the Jensen-Shannon
divergence~\cite{Lin1991} (its square root being a metric).
Figure~\ref{fig:JS_n_grid} shows that with increasing $n$ the Jensen-Shannon
divergence converges to zero.
Here, the finest  phase-space partition has $n=3200^2$ cells.
Furthermore, the ratio of two consecutive Jensen-Shannon divergences is bounded
from above by a value smaller than one.
If this continues to hold for increasing $n$, the coarse-grained measures are a
contractive sequence, therefore a Cauchy sequence and thus converge in the limit
$n \rightarrow \infty$.
This is demonstrated for various decay rates
$\gamma \in [\gamma_{\text{nat}}, \gamma_{\text{inv}}]$ (partial escape) and
$\gamma > \gamma_{\text{nat}}$ (full escape) for all example systems.

Note that for the natural decay rate $\gnat$ (or the inverse natural decay rate
$\ginv$) the construction of the LR-measure $\muLR$ is also possible by a
simpler method:
One can generate the right natural measure by a long-term time evolution of many
phase-space points and their
intensities~\cite{CasMasShe1999b,LeeRimRyuKwoChoKim2004,AltPorTel2013}.
Equivalently, this can be done for the left natural measure by time-reversed
dynamics.
Correspondingly, one finds the LR-measure $\muLR$ for $\gnat$ by iterating
phase-space points forward and backward in time, multiplying the intensities,
and taking the square root.
This was previously done in Ref.~\cite[Fig.~8]{WieMai2008} (without taking the
square root) in order to extend the concept of the chaotic saddle to the case of
partial escape.

\section{Numerical Verification of Factorization}
\label{sec:numerical_verification}

In this section we numerically verify the factorization of the
LR-Husimi representation. First, we explain in~\cref{sec:numerical_factorization}
how we numerically factorize the LR-Husimi representation into an average times
fluctuations.
In~\cref{sec:numerical_fluctuations} we compare the LR-fluctuations
to the theoretically predicted distribution and verify their universality
in the semiclassical limit.
Furthermore, we demonstrate in~\cref{sec:comparison_LR_measure_Husimi}
that the averaged LR-Husimi representation is well described by the LR-measure
and find convergence in the semiclassical limit.

\subsection{Factorization}
\label{sec:numerical_factorization}

For the numerical verification of the factorization of the LR-Husimi
representation, \cref{Eq:factorization_LR_husimi}, we use for the first factor
the averaged LR-Husimi representation, \cref{Eq:averaged_LR_husimi}, giving
\begin{equation}
    \HusLR(\vecx) = \langle \HusLR(\vecx) \rangle_\gamma \cdot
    \etaLR(\vecx)\, .
    \label{Eq:LR_Husimi_decomposition}
\end{equation}
From this we compute the LR-fluctuations by
\begin{equation}
    \etaLR(\vecx) = \frac{\HusLR(\vecx)}{\langle \HusLR(\vecx)
    \rangle_\gamma}\, .
    \label{Eq:LR_fluctuations_numerical}
\end{equation}
Note that for systems with full escape the LR-fluctuations $\etaLR$ should be
evaluated on the chaotic saddle only, as the denominator tends to zero
elsewhere.

Figure \ref{fig:husimi_factorization} shows the factorization of one resonance
state in LR-Husimi representation for all example systems.
For averaging we use 500 resonance states which are close to
the specified decay rate $\gamma$.

Let us mention that the original way of defining the LR-Husimi
representation~\cite{ErmCarSar2009} is without the normalization of
\cref{Eq:LR_husimi_definition},
$\HusLR_n(\vecx) = \left\vert \hLR_n(\vecx) \right\vert$,
such that $\int_\Gamma \ud \vecx\, \HusLR_n(\vecx)$ varies for each
resonance state.
This might affect the averaged LR-Husimi representation and thus the
LR-fluctuations. Numerically, however, we find no significant consequences.

\subsection{LR-fluctuations}
\label{sec:numerical_fluctuations}
\begin{figure*}
	\begin{center}
                \includegraphics[scale=0.85]{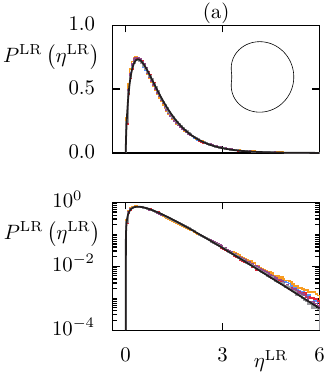}
		\includegraphics[scale=0.85]{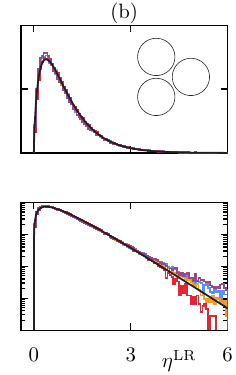}
		\includegraphics[scale=0.85]{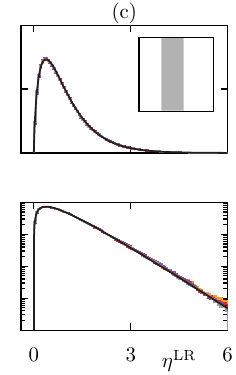}
		\includegraphics[scale=0.85]{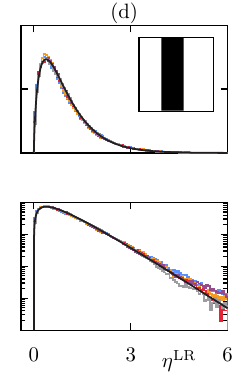}
	\end{center}
        \caption{Visual comparison of the histogram of LR-fluctuations $\etaLR$,
                \cref{Eq:LR_fluctuations_numerical}, to the predicted
                probability distribution $\PLR(\etaLR)$,
                \cref{Eq:LR_fluctuations_distribution} (black), on linear scale
                (top) and logarithmic scale (bottom).
                The resonance states from \cref{fig:husimi_individual} with five
                different decay rates (colored) are used for
                (a) dielectric cavity,
                (b) three-disk scattering system,
                (c) standard map with partial escape, and
                (d) standard map with full escape.
            }
	\label{fig:fluctuations}
\end{figure*}

\begin{figure}
	\begin{center}
		\includegraphics{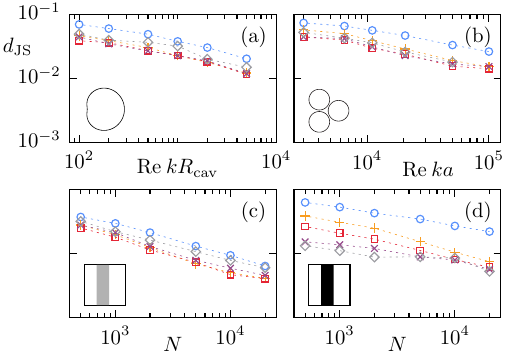}
	\end{center}
	\caption{Convergence of the distribution of LR-fluctuations $\etaLR$,
                \cref{Eq:LR_fluctuations_numerical}, to theoretically predicted
		distribution $\PLR(\etaLR)$, \cref{Eq:LR_fluctuations_distribution},
                for systems and decay rates of~\cref{fig:fluctuations}.
		Shown is the Jensen-Shannon divergence $d_{\text{JS}}$
                (evaluated on 100 bins over the interval $[0, 10]$)
                approaching zero in the semiclassical limit.
	}
	\label{fig:JS_fluctuations}
\end{figure}

We now give numerical evidence that the LR-fluctuations $\etaLR$,
\cref{Eq:LR_fluctuations_numerical}, follow the theoretically predicted
distribution $\PLR(\etaLR)$,~\cref{Eq:LR_fluctuations_distribution}.
This is shown in \cref{fig:fluctuations} for all example systems and
demonstrates the universality of the LR-fluctuations.

In the semiclassical limit
the distribution of the LR-fluctuations $\etaLR$ should converge to
the predicted distribution $\PLR(\etaLR)$
from~\cref{Eq:LR_fluctuations_distribution}.
We verify this
using the Jensen-Shannon divergence~\cite{Lin1991}
as a quantification of the distance of the distributions.
Figure~\ref{fig:JS_fluctuations} shows that going further towards the
semiclassical limit the Jensen-Shannon divergence converges to zero. It is
demonstrated for various decay rates
$\gamma \in [\gamma_{\text{nat}}, \gamma_{\text{inv}}]$ (partial escape) and
$\gamma > \gamma_{\text{nat}}$ (full escape) for all example systems.
A posteriori, this justifies our assumption of statistical independence of the left and right
fluctuations $\etaL$ and $\etaR$ in \cref{sec:distribution_LR_fluctuations}.
In fact, we numerically checked that the Pearson correlation coefficient between
the left and right fluctuations $\etaL$ and $\etaR$ decays in the semiclassical
limit (not shown).

These results on the distribution of LR-fluctuations explain some of the
findings of Ref.~\cite{MonCarBor2024}.
In particular, the comparison of the LR-fluctuations to an exponential
distribution shows strong deviations, as shown in
Ref.~\cite[Fig.~1]{MonCarBor2024}.
This observation is consistent with the above prediction that the
LR-fluctuations are distributed according to
\cref{Eq:LR_fluctuations_distribution}.
Furthermore, in Ref.~\cite[Fig.~2]{MonCarBor2024} a localization measure shows
differences between the LR-fluctuations and the fluctuations of the Husimi
representation of right resonance states.
We attribute this to the different underlying universal distributions, namely
\cref{Eq:LR_fluctuations_distribution} compared to the exponential distribution.

\subsection{Averaged LR-Husimi representation}
\label{sec:comparison_LR_measure_Husimi}

\begin{figure*}
	\begin{center}
                \includegraphics[width=1\linewidth]{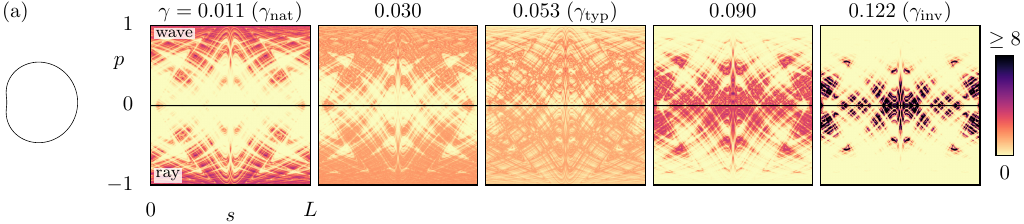}

		\vspace{0.5cm}
                \includegraphics[width=1\linewidth]{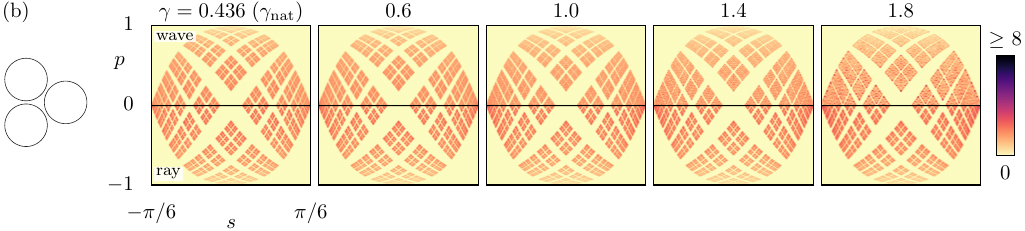}

		\vspace{0.5cm}
                \includegraphics[width=1\linewidth]{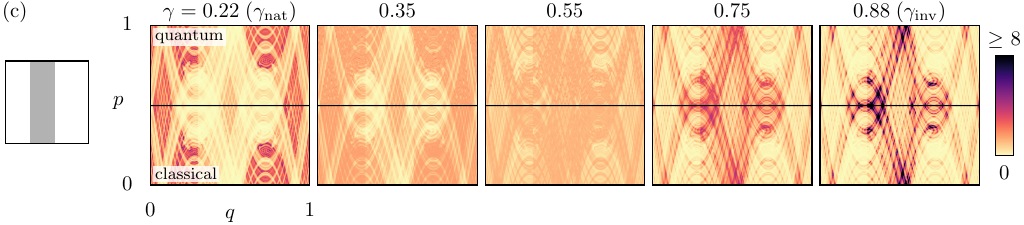}

		\vspace{0.5cm}
                \includegraphics[width=1\linewidth]{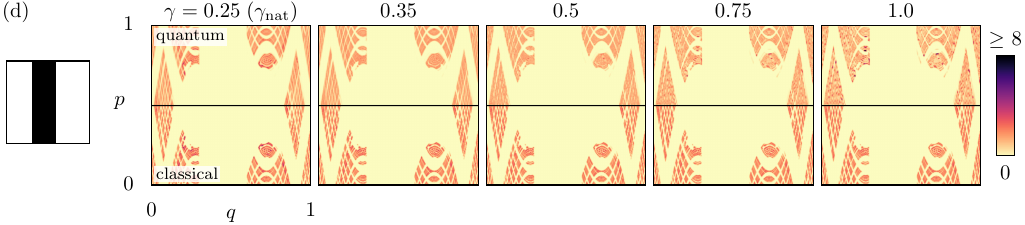}
	\end{center}
        \caption{Visual comparison of the averaged LR-Husimi representation $\langle \HusLR
                \rangle_\gamma$ (upper half) to the proposed LR-measure $\muLR$
                smoothed on the scale of a Planck cell
                (lower half) for five different decay rates.
                For the average we use 500 resonance states close to the given decay
                rate $\gamma$.
                This is shown for
                (a) dielectric cavity,
                (b) three-disk scattering system,
                (c) standard map with partial escape, and
                (d) standard map with full escape.
	}
	\label{fig:husimi_ray_wave_comparison}
\end{figure*}

\begin{figure}
	\begin{center}
		\includegraphics{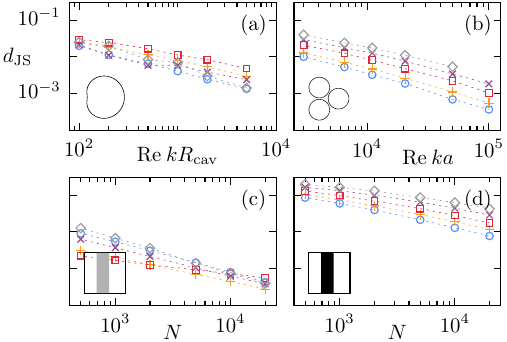}
	\end{center}
	\caption{Convergence of averaged LR-Husimi representation $\HusLRAvg$
                to LR-measure $\muLR$ for systems and
		decay rates of~\cref{fig:husimi_ray_wave_comparison}.
                Shown is the Jensen-Shannon divergence
                $d_{\text{JS}}$ (evaluated on a $50 \times 50$
                grid) approaching zero in the semiclassical limit.
	}
	\label{fig:JS_ray_wave_avg}
\end{figure}

In the following we show that the averaged LR-Husimi representation
$\langle \HusLR \rangle_\gamma$ is described by the LR-measure $\muLR$.
To this end we first give a visual comparison for all example systems and
various decay rates $\gamma$ in~\cref{fig:husimi_ray_wave_comparison}.
Here the LR-measure $\muLR$ is computed on a $3200 \times 3200$ grid
($n \approx 10^7$ partition cells).
For visual comparison we smooth it on the scale corresponding to the Planck cell $h$
of the LR-Husimi representation.
We see an overall good agreement between the averaged LR-Husimi representation
and the LR-measure for all cases.
Small differences appear only on the scale of a Planck cell, see e.g.\ $\gtyp$
(middle) in \cref{fig:husimi_ray_wave_comparison}(a) and the largest decay rate
in \cref{fig:husimi_ray_wave_comparison}(b).

An even better visual comparison can be achieved by first smoothing the
individual left and right measure before their multiplication
in~\cref{Eq:LR_measure}, see Appendix~\ref{sec:appendix_alternative_smoothing}.
However, this is not necessary in order to find agreement
in the semiclassical limit, see below.

For systems with full escape we observe that the LR-measure
does not depend on $\gamma$ and that it is uniform
on the chaotic saddle.
We explain this using the fact that the dynamical system confined to
the chaotic saddle is closed and ergodic implying a uniform distribution
of the LR-measure.
Note that for the three-disk scattering system the LR-measure is multiplied by
$\sqrt{1-p^2}$ for the visual comparison as this factor is a property of the
boundary Husimi representation~\cite{BaeFueSch2004}.

In the semiclassical limit
the averaged LR-Husimi representation $\HusLRAvg$ should converge to the
LR-measure $\muLR$.
We verify this using the Jensen-Shannon divergence
as a quantification of the distance between $\HusLRAvg$ and $\muLR$.
Figure~\ref{fig:JS_ray_wave_avg} shows that going further towards the
semiclassical limit the Jensen-Shannon divergence converges to zero.
This is demonstrated for various decay rates
$\gamma \in [\gamma_{\text{nat}}, \gamma_{\text{inv}}]$ (partial escape) and
$\gamma > \gamma_{\text{nat}}$ (full escape) for all example systems.

\section{Summary and Outlook}
\label{sec:discussion}

In summary, we demonstrate the quantum invariance of the LR-Husimi
representation and show that it can be factorized into a product of a smooth
classical density $\rhoLR$ and universal LR-fluctuations $\etaLR$.
Quantum invariance is introduced as a property of resonance states
in their LR-Husimi representation and states that in the semiclassical limit
$h \to 0$ and for a phase-space region with fixed size it is invariant under the
closed dynamics.
In the factorization of the LR-Husimi representation the first factor $\rhoLR$
is a classical density, smooth on the scale of a Planck cell,
determined from the conditionally
invariant measures of the left and right resonance states.
Its structure strongly depends on the specific system.
For systems with partial escape there is also a strong dependence on the decay
rate.
In contrast, for systems with full escape, $\rhoLR$ is uniform on the
chaotic saddle and is independent of $\gamma$.
The second factor $\etaLR$ describes fluctuations around this average and
follows a universal distribution, \cref{Eq:LR_fluctuations_distribution}, that
is independent of the system and the decay rate.
The derivation of this distribution is based on the assumption of statistical
independence of the left and right fluctuations.
We justify this assumption a posteriori by the numerical verification of the
distribution.

We expect that the coarse-grained LR-measure fulfills invariance under the
closed dynamics similar to the quantum invariance of the LR-Husimi
representation.
Here, the finite-sized phase-space partition of Ulam's method replaces the
finite-sized Planck cell.
In the limit of an infinitely fine partition this should lead to an invariant
measure $\muLR$ under the closed dynamics, which could be tested numerically.
If so, then for systems with partial escape one finds a one-parameter family of
invariant measures depending on the decay rate $\gamma$.
It is an open question whether or not these are related to the invariant
measures (equilibrium states) of the thermodynamic formalism in ergodic theory,
which depend on the inverse temperature $\beta$~\cite{Wal1982, BecSch1993,
Rue2004}.

Furthermore, it is an interesting open problem whether the presented results
on quantum invariance for
fully chaotic system generalize to systems with a mixed (regular and chaotic)
phase space.
The analytical arguments in \cref{sec:invariance}
do not depend on the type of dynamics and thus seem to suggest that quantum
invariance holds more generally for resonance states with a well defined
semiclassical limit.

An important open question is how the presented results can be observed
in experimental scattering systems.
As we are in a regime of strongly overlapping resonances, the properties of an
individual resonance are experimentally not accessible.
However, the average over many resonances with similar decay rates also depends
on the decay rate and might be observable by the time evolution of wave packets.
This is experimentally related to pulse propagation deduced from transmission
spectra~\cite{SchKuhSto2006}.

\section*{Data availability statement}
The data that support the findings of this study are openly available at the
following URL/DOI: \url{https://doi.org/10.25532/OPARA-989}.

\acknowledgments

We are grateful for discussions with A.~Bäcker, E.-M.~Graefe, and D.~Savin.
RK thanks M.~M.~Castro and G.~Froyland for illuminating discussions on invariant
measures and the possible relation of this work to equilibrium states of ergodic
theory.
We thank L.~M\"uller for contributions in the early stage of the project~\cite{Mue2024}.
Funded by the Deutsche Forschungsgemeinschaft (DFG, German Research
Foundation) -- 262765445.
The authors gratefully acknowledge the computing time on
the high-performance computer at the NHR Center of TU Dresden.
This center is jointly supported by the Federal Ministry of Education and
Research and the state governments participating in the NHR.


\begin{figure*}[b]
	\begin{center}
		\includegraphics[width=1\linewidth]{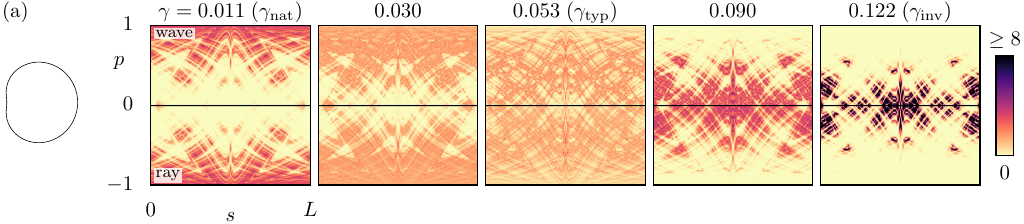}

		\vspace{0.5cm}
		\includegraphics[width=1\linewidth]{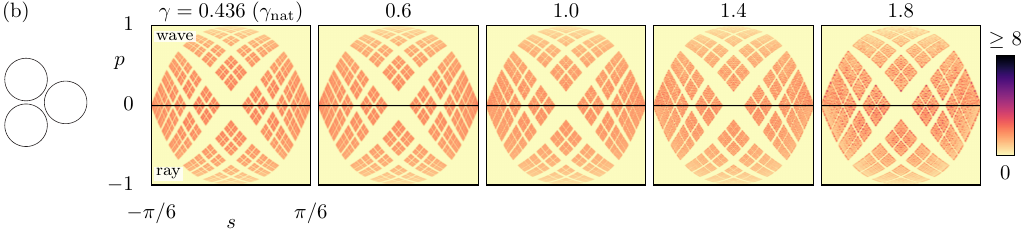}

		\vspace{0.5cm}
		\includegraphics[width=1\linewidth]{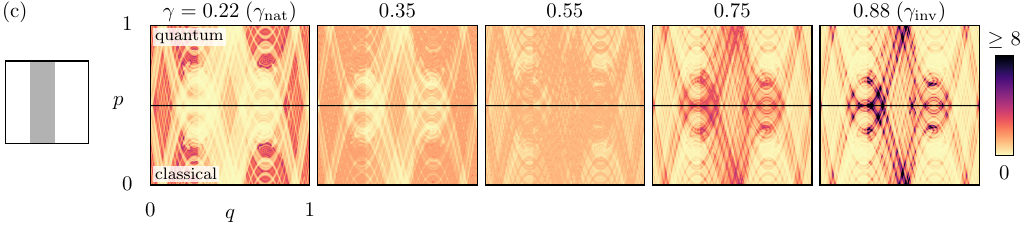}

		\vspace{0.5cm}
		\includegraphics[width=1\linewidth]{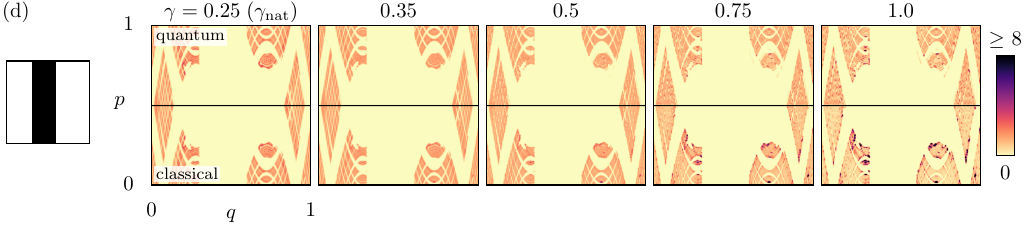}
	\end{center}
        \caption{Visual comparison of the averaged LR-Husimi representation
                $\langle \HusLR \rangle_\gamma$ (upper half, same as in
                \cref{fig:husimi_ray_wave_comparison}) to an alternative LR-measure
                $\muLR$ with smoothing on the scale of a Planck cell before multiplication
                (lower half) for five different decay rates.
                This is shown for
                (a) dielectric cavity,
                (b) three-disk scattering system,
                (c) standard map with partial escape, and
                (d) standard map with full escape.
	}
	\label{fig:husimi_ray_wave_comparison_appendix}
\end{figure*}

\appendix

\section{Alternative smoothing of LR-measure}
\label{sec:appendix_alternative_smoothing}

The comparison of the averaged LR-Husimi
representation $\HusLRAvg$ to the LR-measure $\muLR$ in \cref{fig:JS_ray_wave_avg}
convergences in the semiclassical limit, but the visual comparison for finite
wavelengths in \cref{fig:husimi_ray_wave_comparison} shows deviations on the
scale of a Planck cell.
This is improved by smoothing the individual left and right measures $\muL$ and
$\muR$ on the scale of a Planck cell before multiplying them according to
\cref{Eq:LR_measure}, see \cref{fig:husimi_ray_wave_comparison_appendix}.
We motivate this by the fact that the individual left and right Husimi
representations are also smooth on the scale of a Planck cell before being
multiplied in \cref{Eq:LR_husimi} to obtain the LR-Husimi representation.

At finite wavelengths this alternative construction of the LR-measure is
better able to reproduce the intensity maxima in the LR-Husimi representation,
e.g.\ compare Figs.~\ref{fig:husimi_ray_wave_comparison_appendix}(a) and
\ref{fig:husimi_ray_wave_comparison}(a) at $\gtyp$ (middle) or
Figs.~\ref{fig:husimi_ray_wave_comparison_appendix}(b) and
\ref{fig:husimi_ray_wave_comparison}(b) at the largest decay rate.
Additionally, for full escape the smallest scale of the multifractal structure
is better reproduced in \cref{fig:husimi_ray_wave_comparison_appendix}(b,~d)
than in \cref{fig:husimi_ray_wave_comparison}(b,~d) for all decay rates.

However, we also observe a drawback of this alternative construction.
For the three-disk scattering system,
\cref{fig:husimi_ray_wave_comparison_appendix}(b), one observes a
dependence of the measure $\muLR$ on the $s$-direction for $p \approx 0$ for
larger decay rates.
This does not occur for the averaged LR-Husimi representation.
Additionally, it has the disadvantage that the construction of the LR-measure
depends on the size of the Planck cell.
This is why we prefer the construction of the LR-measure $\muLR$ in the main
text, \cref{Eq:LR_measure}, with smoothing just for visualization purposes.


\end{document}